\input{aipcheck.tex}

\documentclass{aipproc}

\layoutstyle{8x11double}

\SetInternalRegister\hbadness{8000}

\begin{document}

\title 
      [Spectral Analysis of Bright Gamma Ray Bursts]
      {Spectral Analysis of Bright Gamma Ray Bursts}


\author{G. Ghirlanda}{
  address={SISSA/ISAS, Trieste - Via Beirut 2, 34014 Trieste, Italy},
  email={ghirland@sissa.it}
}

\iftrue
\author{A. Celotti}{
  address={SISSA/ISAS, Trieste - Via Beirut 2, 34014 Trieste, Italy}
}

\author{G. Ghisellini}{
  address={Osservatorio Astronomico di Brera, via Bianchi 46, 
Merate (LC), Italy}
}
\fi

\copyrightyear  {2001}

\begin{abstract}
We present the time
integrated and time resolved spectral analysis of a sample
of bright bursts selected from the BATSE archive. 
We fitted four different spectral models to the
time integrated and time resolved spectra of the flux pulses.
We point out that the found (marginal) differences in the parameter
distributions can be ascribed to the different spectral shape of the employed models
and that a smoothly curved model best fits the observed spectra.
We characterize the spectral shape of bright bursts and compare the low
energy slope of the fitted spectra with the prediction $N(E)\propto E^{-2/3}$ of the
synchrotron theory, finding that this limit is violated in a considerable
number of time resolved spectra around the peaks,
both during the rise and decay phase.
\end{abstract}

\date{\today}

\maketitle

\section{Introduction}
The nature and emission mechanisms responsible for the
prompt emission of Gamma--ray bursts (GRB) are still unclear.
In order to identify the physical process(es)
responsible for the emission it would be ideally necessary
to study spectra resolved on the shortest time--scales of
variability, typically of a few milliseconds, observed in
bright bursts \cite{Fishman} and
predicted on theoretical basis \cite{Rees}.
In fact time resolved spectra, even within
single pulses, show a strong time evolution, and are, in
general, harder than time integrated spectra
\cite{Liang}, \cite{Crider a}.

Here we present the study of the
spectral properties of single pulses within bright GRBs,
compare the results of the spectral analysis of the time average
spectrum with the time resolved spectra of the very same burst in
order to quantify systematic differences and examine any spectral
`violation' (with respect to the predicted slope in the case of e.g. 
synchrotron
emission) for the entire burst evolution.
A more detailed analysis is given in \cite{Ghirlanda}.

\section{Data Analysis}
The main characteristics of the BATSE detectors have been described by
\cite{Fishman}.  We selected bursts with a peak flux, on the 64 ms
time-scale, higher than $20\ \mathrm{phot/cm^{2} sec}$.  The data used
were mainly the HERB: time sequence of 128 channel spectra with a
minimum integration time of 0.128 s.

Within the selected bursts each peak was analyzed separately
considering the spectrum time-integrated over the duration of the peak
and the sequence of time resolved spectra comprised within the same
peak.  We fitted the background subtracted spectra with 4 spectral
models: the BAND one (Band et al.
\cite{Band}) which consists of 2 power laws joined smoothly by an exponential 
roll--over,
the Broken Power Law (BPLW hereafter) which has a sharp break between
the two power law segments, the COMP model which comprises a low
energy power law ending--up in an exponential cutoff, and the
Synchrotron Shock Model (SSM) \cite{Tavani} based on optically thin
synchrotron emission from relativistic particles (and for the first
time fitted to the time resolved spectra).

\section{Results}

From a statistical point of view,
the comparison of the spectral fits obtained with the 4 models shows
that in terms of the reduced $\chi^{2}$, 
the BAND and COMP models can better represent the time resolved
spectra of bright bursts, but some counter examples exist showing that
in general within a single pulse more than one time resolved spectrum
can be fitted by different spectral models.  


\subsection{The parameters distributions}
\begin{figure}
\resizebox{\columnwidth}{!}
  {\includegraphics{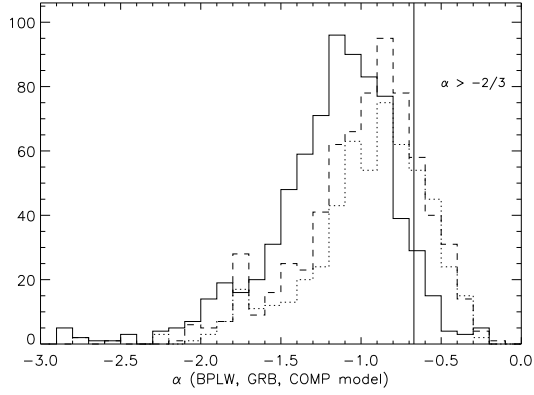}}
  \caption{Low energy power law spectral index ($\alpha$)
      distributions derived from the time resolved spectral
      analysis. \textit{Solid line:} BPLW model, \textit{dotted line:}
      BAND model, \textit{dashed line:} COMP model. The vertical line
      represents the synchrotron limit ($\alpha=-2/3$) for the low
      energy spectral shape.}   \label{fig:a}
\end{figure}

The distributions of the \emph{low energy power law spectral index}
$\alpha$ (Fig.\ref{fig:a}), for the BAND and COMP model, are similar,
like in the case of the time integrated spectra, and both have a mode
of $-0.85\pm0.1$, also consistent with the BPLW average value $-1.15
\pm 0.1$. Note that qualitatively the extension of the $\alpha$
distribution of the BPLW model (\textit{solid line} in
Fig.\ref{fig:a}) towards lower values could be attributed to the fact
that at low energies this model (which has a sharp break) tends to
under--estimate the hardness of the spectrum compared to a smoothly
curved model.

The average low energy spectral slope obtained from the time resolved
spectra is harder than what obtained with the time integrated pulse
spectra for all the three models (BAND, BPLW, COMP). This is a
consequence of time integration (i.e. hardness averaging) of the
spectral evolution (which can be also very dramatic) over the entire
rise and decay phase of the pulse.

\begin{figure}
\resizebox{\columnwidth}{!}
  {\includegraphics{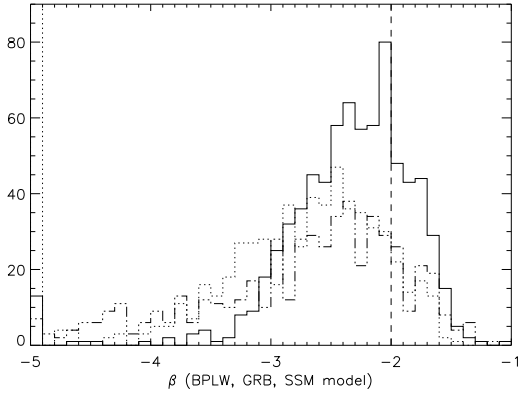}} \caption{High energy power law
  spectral index ($\beta$) distributions for the time resolved
  spectra. \textit{Solid line:} BPLW model; \textit{dotted line:} BAND
  model; \textit{dot--dashed line:} SSM model. Also shown (bin
  with $\beta=-5$) the time resolved spectra with undetermined high
  energy spectral index for the BAND model.} \label{fig:b}
\end{figure}

The \emph{high energy spectral index} $\beta$ distributions
are reported in Fig.\ref{fig:b}.
The average value is $-2.45 \pm 0.1$ and
$-2.05 \pm 0.1$ for the BAND and BPLW model respectively, the former being harder
than what obtained from the pulse average spectrum.
The SSM average $\beta$ is -2.17 which is consistent with
what found from the average pulse spectral analysis.

\begin{figure}
\resizebox{\columnwidth}{!}
  {\includegraphics{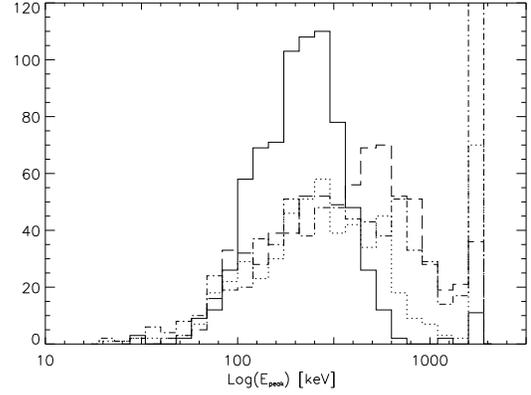}}
  \caption{$E_{peak}$. Peak energy distribution for the 4 spectral models.
\textit{solid line:}BPLW model,\textit{dotted line:}BAND model,
\textit{dashed line:}COMP model,\textit{dot--dashed line:}SSM model. Spectra with undetermined
peak energy (i.e. the high energy threshold 1800 keV assumed as lower limit) are reported in
the last bin.}   \label{fig:c}
\end{figure}

The most important spectral parameter obtained in fitting
the spectrum with these models is $E_{peak}$ corresponding
to the peak of the $E F_E$ spectrum, and thus to the energy
where most of the power is released.
$E_{peak}$ is coincident with the break energy
$E_{0}$ for the BPLW and COMP and is equal to $(\alpha +
2)E_{0}$ for the BAND model. This
characteristic energy can be obviously calculated only for
those spectra (BPLW and BAND model) with $\beta < -2$ and its distribution
is presented
in Fig.\ref{fig:c}.
The average is $E_{peak} = 280^{+72}_{-57}$ keV for the BAND model,
consistent, within the errors, with the BPLW most
probable value of $211^{+25}_{-22}$ keV. The COMP model,
instead, gives a highly asymmetric peak energy distribution
with a mode of $595^{+104}_{-88}$ keV because the lack of
an high energy power law component tends to over--estimate
the energy corresponding to the start of the exponential
cutoff. The SSM model has an average $E_{peak} \sim 316^{+64}_{-52}$ keV
with a wide distribution.

\subsection{The Synchrotron limit violation}

A strong prediction of the optically thin
synchrotron model is that the asymptotic low energy photon
slope $\alpha$ should be lower than or equal to $-2/3$ \cite{Katz}.

\begin{figure}
 \resizebox{\columnwidth}{!}  {\includegraphics{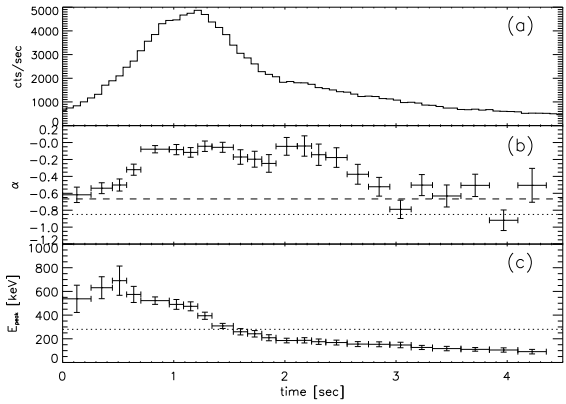}}
  \caption{Trigger 2083. Spectral evolution of the BAND model fitted
  to the time resolved spectra. Light curve on the 64 ms time-scale
  (panel a); low energy spectral index and (\textit{dashed})
  synchrotron shock model limit $\alpha=-2/3$ (b); peak energy
  (c). For reference the average values of $\alpha$ and $E_{peak}$
  obtained from the time resolved spectra (\textit{dotted line}) and
  the synchrotron model limit are reported (\textit{dashed line}).}
\label{fig:d}
\end{figure}

\begin{figure}
\resizebox{\columnwidth}{!}
  {\includegraphics{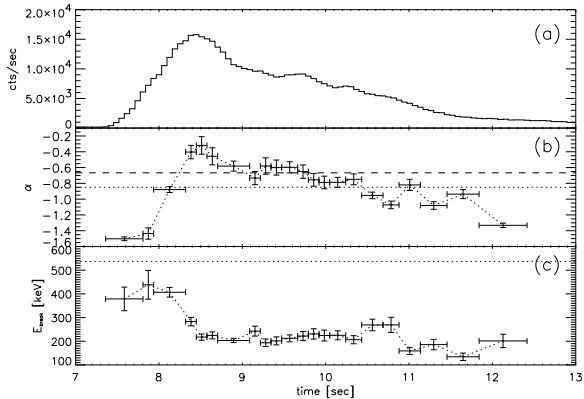}} \caption{Trigger 5614. Spectral
  evolution of the COMP model fitted to the time resolved
  spectra. Light curve on the 64 ms time--scale (panel a); low energy
  spectral index and (\textit{dashed}) synchrotron shock model limit
  $\alpha=-2/3$ (b); peak energy (c). For reference the average values
  of $\alpha$ and $E_{peak}$ obtained from the time resolved spectra
  (\textit{dotted line}) and the synchrotron model limit
  (\textit{dashed line}) are reported.}
\label{fig:e}
\end{figure}

We obtain that the 13.7\% of the time resolved spectra fitted with the
BAND model are inconsistent with $\alpha \le -2/3$ at 2$\sigma$.  A
similar percentage of spectra violating the $\alpha$ limit is found
for the COMP model ($\sim$11.7\%, of course mostly for the same
spectra). Moreover $\sim$ 21\% of the time resolved peak spectra
violate the synchrotron limit, indicating that this violation happens
during the peak phase and not preferentially before or after it: in
Fig.\ref{fig:d} we show the spectral evolution in the case of
GRB921207 (fitted with the BAND model) and another example is reported
in Fig.\ref{fig:e}.

\section{Conclusions}
We considered a sample of bright burst detected by BATSE and performed
a uniform analysis for the time integrated and the time resolved
(typically 128 ms) spectra with four different models adopted and
proposed in the literature.

We find that even at this time resolution no model can better
represent the data and different spectra require different shapes,
re-confirming the erratic behavior of bursts and also possibly
indicating that time resolution on time-scales comparable with the
variability one are needed.  The parameter distributions are
consistent with the results reported by \cite{Preece a} although the
average spectral shape (both from the time resolved and time
integrated spectra) is harder because we selected only bright bursts
and restricted the spectral analysis to the pulse phase.

A considerable number of spectra are characterized by extremely
hard low energy components with spectral index greater than --2/3
(i.e. the limit predicted by synchrotron theory \cite{Katz}).  The
$\alpha$ limit violation, also found by \cite{Frontera} and
\cite{Crider a}, is evident around the peak both during the rise and
decay phase, and this could indicate that at some stages of the burst
evolution (possibly near the peak of emission itself) alternative
radiative processes, other than synchrotron, can be dominant.
%
\begin{theacknowledgments}
This research has made use of data obtained through the High Energy
Astrophysics Science Archive Research Center Online Service, provided
by the NASA/Goddard Space Flight Center.  We are grateful to D. Band
for useful discussions on the BATSE data analysis. We thank Marco
Tavani for having provided his code of the SSM model. Giancarlo
Ghirlanda and AC acknowledge the Italian MUIR for financial support.
\end{theacknowledgments} 

%

\end{document}